\documentclass[sigconf]{acmart}

\AtBeginDocument{%
  \providecommand\BibTeX{{%
    \normalfont B\kern-0.5em{\scshape i\kern-0.25em b}\kern-0.8em\TeX}}}

\copyrightyear{2023}
\acmYear{2023}
\setcopyright{acmlicensed}
\acmConference[AIES '23]{AAAI/ACM Conference on AI, Ethics, and Society}{August 8--10, 2023}{Montréal, QC, Canada}
\acmBooktitle{AAAI/ACM Conference on AI, Ethics, and Society (AIES '23), August 8--10, 2023, Montréal, QC, Canada}
\acmPrice{15.00}
\acmDOI{10.1145/3600211.3604686}
\acmISBN{979-8-4007-0231-0/23/08}

\newcounter{daggerfootnote}
\newcommand*{\daggerfootnote}[1]{%
    \setcounter{daggerfootnote}{\value{footnote}}%
    \renewcommand*{\thefootnote}{\fnsymbol{footnote}}%
    \footnote[2]{#1}%
    \setcounter{footnote}{\value{daggerfootnote}}%
    \renewcommand*{\thefootnote}{\arabic{footnote}}%
    }

\usepackage{enumerate}

\begin{document}

\title{The Ethical Implications of Generative Audio Models: A Systematic Literature Review}

\author{Julia Barnett}
\email{JuliaBarnett@u.northwestern.edu}
\orcid{0000-0002-3476-1110}
\affiliation{%
  \institution{Northwestern University}
  \streetaddress{}
  \city{Evanston}
  \state{IL}
  \country{USA}
}

\renewcommand{\shortauthors}{Julia Barnett}

\begin{abstract}
  Generative audio models typically focus their applications in music and speech generation, with recent models having human-like quality in their audio output. This paper conducts a systematic literature review of 884 papers in the area of generative audio models in order to both quantify the degree to which researchers in the field are considering potential negative impacts and identify the types of ethical implications researchers in this area need to consider. Though 65\% of generative audio research papers note positive potential impacts of their work, less than 10\% discuss any negative impacts. This jarringly small percentage of papers considering negative impact is particularly worrying because the issues brought to light by the few papers doing so are raising serious ethical implications and concerns relevant to the broader field such as the potential for fraud, deep-fakes, and copyright infringement. By quantifying this lack of ethical consideration in generative audio research and identifying key areas of potential harm, this paper lays the groundwork for future work in the field at a critical point in time in order to guide more conscientious research as this field progresses.
\end{abstract}



\maketitle

\section{Introduction}

Generative models have been a large focus of AI researchers over the past few years, and recently the public has seen these models first-hand in public facing algorithms like ChatGPT \cite{openai_2022} for text, DALLE-2 \cite{ramesh2022hierarchical} for vision, and Jukebox \cite{dhariwal2020jukebox}\protect\daggerfootnote{References for works included and analyzed in the systematic literature review corpus can be found in the Appendix. They are denoted in text with the dagger$^\dag$ superscript.} for music.  At their core, generative models are a type of AI system that take in vast amounts of training data to be able to produce a novel item that is similar to and statistically likely to exist in the data it was trained on. Though generative models have been around for decades with origins in the 1980s \cite{bond2021deep}, the outputs of these models saw unprecedented advances with the introduction of the transformer in 2017 which revolutionized the field by introducing a mechanism called ``attention'' that allowed for much more accurate and complex outputs of generative models \cite{vaswani2017attention}. Generative models may continue to improve as (a) their training data becomes larger (for text, imagine the entire internet) and (b) researchers continue to make advances in the architecture of the models. This paper focuses specifically on the current landscape of generative audio models. 

As generative audio models continue to develop and grow both in popularity and complexity, this research seeks to understand the ethical landscape of potential impacts of these models. In particular, this paper explores what potential harms have been considered by researchers creating deep generative modeling projects, and seeks to understand the extent to which researchers in this domain are considering the broader ethical implications of their work. When a layperson is introduced to generative models their instinct is to jump to potential negative impacts \cite{edelman_survey, kelley2021exciting}, however, researchers in the field are wary to do the same. There has been minimal research into the ethical implications of deep generative audio models, and this paper calls out the need for that to change by providing a comprehensive and thorough overview of the current potential negative impact domain.

Systematic literature reviews are effective at evaluating the current landscape of a research domain---especially when the potential corpus to analyze is a tractable number. In addition to identifying trends, they are particularly helpful in identifying gaps in the field. This is an agenda setting paper at the right time---it is important to both diagnose the degree to which research papers on generative audio models are discussing ethics and encourage the plethora of researchers to come to include a negative broader impact in their analysis prior to the field being clogged by studies without an ethical component. As will be discussed in more detail below in Section \ref{sec:gen_audio_models}, innovations in text and vision typically precede those in generative audio, so this same analysis conducted in the generative text and vision domains would include 3,099 and 5,287 articles, respectively. 884 papers in the generative audio domain prior to screening is comparably a more tractable endeavor to undertake, and it is feasible to raise the concern of lack of negative impacts earlier in this specific area.

This paper makes two concrete contributions. Primarily, it quantifies the degree to which researchers in the generative audio domain consider the ethical implications and negative broader impacts of their work. The author finds that less than 10\% of the corpus ($n=16/171$) discuss any potential negative impact. Secondarily, this paper examines the different negative broader impacts explored by these 16 papers and thematically discusses the potential ethical implications. This paper calls to attention the need for researchers in this domain to consider the ethical implications of their work, and suggests a starting point for topics to consider by examining issues already brought to light by their peers.

\section{Background and Definitions}

\subsection{Generative Audio Models}\label{sec:gen_audio_models}

Generative models can largely be grouped into three buckets: text, vision, and audio. Due to the nature of the data underneath the models, advances typically start in the textual domain, followed by images, and finally audio. The most clear-cut example of this is when Vaswani et al. introduced the first transformer for text in 2017 \cite{vaswani2017attention}, which led to the first image transformer in June of 2018 \cite{imagetransformer2018}, and shortly thereafter, the music transformer \cite{huang2018music}. At their core, generative models use a large amount of training data in order to predict something that is similar to and statistically likely to exist in the dataset it was trained on. Compared to text generation models, which are limited by the finite vocabulary of the language being used, audio can have exponentially larger potential combinations of sound occurring at once; even if just isolating to the possible keys of a piano, the possibilities become almost intractably large extremely fast in the granularity of milliseconds.

Despite these data challenges, many advances have been made in the generative audio domain mainly in the areas of music and speech generation. Music generation is particularly tricky due to the long-term relational dependencies of melodies and other musical structure that may have occurred at a timestamp far earlier than the current frame needing to be generated \cite{huang2018music}, but self-attention and relative positioning \cite{shaw-etal-2018-self} enabled that hurdle to be overcome. Generative models today can condition on artist and genre to steer the style of music \cite{dhariwal2020jukebox}$^\dag$, or even condition on text and melodies to create high-fidelity and quality musical compositions \cite{agostinelli2023musiclm}$^\dag$. Advances in speech have varied from speech enhancement and denoising \cite{polyak2021high, zhang2021restoring}$^\dag$, text-to-speech (TTS) generation \cite{habib2019semi, kim2021conditional}$^\dag$, accent conversion and style transfer \cite{polyak2020unsupervised, zhou2018voice}$^\dag$, and audio in-painting to reconstruct gaps in speech data \cite{tae2021editts}$^\dag$. Most of the research in generative audio models is concentrated in music and speech generation; however, there are some cases where the models generate specifically non-music, non-speech sounds such as auxiliary sound effects for movies or birds chirping \cite{ghose2020autofoley, zhuo2023exploring}.

Though audio generation can also be tied together with visual generation in the forms of videos and deepfakes \cite{diakopoulos2021anticipating, Fried2019talkinghead, meskys2020regulating}, motion to create dance moves and choreography \cite{au2022choreograph, Valle2021Transflower, Zhuang2022dance} or lip movements and other speech gestures \cite{Jonell2020lets, waibel2022face, yoon2020Gesture}, this work focuses on generated audio in the singular medium. For example, while text-to-speech works will be evaluated, speech-to-text work will not, nor will creating a dance routine simultaneously with music. The goal of this isolation of audio is to understand explicitly the ethical discussions of the audio domain, not to potentially conflate these issues with ethical discussions of other fields.

\subsection{Broader Impact}\label{sec:broader_impact}

Especially with the public spotlight on deep generative models such as ChatGPT \cite{openai_2022}, both computer scientists and the public alike have become aware of the potential negative impacts of these models and other algorithmic systems. A recent thematic review of broader impact statements of the Neural Information Processing Systems (NeurIPs) 2020 conference found that some of these broader categories related both to how consequences are expressed such as specificity and uncertainty as well as different areas of impacts expressed such as bias, the environment, labor, and privacy \cite{nanayakkara2021unpacking}. A recent survey of the socio-technical harms of algorithmic systems identified five major types of harms: representational, allocative, quality-of-service, interpersonal, and societal harms  \cite{shelby2022sociotechnical} in order to establish conceptual alignment for future research and to encourage consideration of these negative impacts and reduce the harm these systems cause.

There are a variety of approaches to encourage broader impact consideration in scientific research. The US National Science Foundation and other grant providers require a Broader Impacts Criterion in both grant applications and  the peer review process \cite{mardis2012toward}, though there is mixed reception around this being the best manner to encourage consideration of societal impact \cite{holbrook2011peer, roberts2009realizing}. Other researchers in computer science have suggested that a simple change to the peer review process would substantially change the degree to which computer scientists consider the negative impact of their work \cite{hecht2021s}. 

Another method recently proposed utilizes crowdsourcing to anticipate different societal impacts of algorithmic decision making systems \cite{barnett2022crowdsourcing}, which puts the consideration in the hands of the layperson in addition to the algorithm designer in order to have a comprehensive idea of potential impacts. Other methods include impact assessment tools such as algorithmic impact assessments (AIAs) which strive to both identify varying areas of impact in addition to establishing steps to hold the algorithm creators accountable \cite{cashmore2009introduction, metcalf2021algorithmic}. Ethics and society review (ESR) was a recently piloted program that facilitated ethical and societal reflection as a requirement to secure funding. They found that 100\% of participants saw the benefit in the process and were willing to continue submitting projects in this manner \cite{bernstein2021ethics} indicating that the demand is there among researchers to consider ethical impact. 

The uniting thread of all of these methods is to encourage societal impact beyond the main text of the paper, or even to require a third party to assist in the ethical evaluation. This paper instead focuses on research papers themselves (as opposed to a secondary document/evaluation such as a grant proposal or peer review) and the extent to which they consider broader impact in the main body and appendices.

Prior research by Weidinger et al. has established a taxonomy of ethical and social risks of harm from language models \cite{weidinger2021ethical}, which in this paper is extended to generative audio models and helps guide the definition of broader impact. Weidinger et al. classify six areas of harms of language models: (1) discrimination, exclusion, and toxicity, (2) information hazards, (3) misinformation harms, (4) malicious uses, (5) human-computer interaction harms, and (6) automation, access, and environmental harms. Discrimination, exclusion, and toxicity focuses on the different treatment of social groups in an oppressive manner. Information hazards concern privacy violations and safety risks, such as compromising privacy due to systems that leak or enable the correct inference of private information. Misinformation harms have to do with the dissemination of misleading information leading to material harm, for instance in the cases of medical misinformation leading to serious consequences for people's quality of life \cite{swire2019public}. Malicious uses are explored more broadly for AI systems by Brundage et al., and they define these as ``all practices that are intended to compromise the security of individuals, groups, or a society'' \cite{brundage2018malicious}. Human-computer interaction harms encompass harms from the direct interaction of humans with the AI system. Finally, automation, access, and environmental harms highlight downstream application impacts that benefit access to select groups and not society at large.

For the purposes of this paper, and guided by Weidinger et al's taxonomy detailed above, broader impact is defined as a possible impact or application of the research/model on the broader society, rather than the scoped technological or scientific purposes. For example, the explicit purpose and scientifically relevant impact of a music generative model is to create music, possibly with long-term structure \cite{huang2018music} or guided by text inputs \cite{agostinelli2023musiclm}$^\dag$. A positive broader impact in this case could be to creatively inspire musicians, and a negative impact could be copyright violations. This analysis will focus primarily on the extent to which negative impacts are discussed and explored, but will also note when researchers discuss positive broader impacts beyond their scientific scope.

\subsection{Research Questions}

The formal research questions addressed by this paper are:
\begin{enumerate}
    \item To what extent is the current study of generative audio models addressing negative broader impacts?
    \item What ethical considerations of generative audio models has the field examined?
\end{enumerate}

\section{Data and Methodology}

In order to address these two research questions, the author conducted a systematic literature review (SLR) of research articles published over the last five years in the generative audio domain. The reporting of this SLR was guided by the standards of the Preferred Reporting Items for Systematic Reviews and Meta-Analyses (PRISMA) guidelines \cite{moher2009preferred} in order to transparently and concisely evaluate the current state of the field. This was an ideal methodology for these research questions due to the nature of the study evaluating what the field has done in a comprehensive and broad sense.

\subsection{Search Strategy}

\subsubsection{Inclusion and Exclusion Criteria}\label{sec:inclusion_criteria}

Formatively, this study analyzed full research papers in the generative audio domain. This does not include extended abstracts or book chapters, nor any other form of writing outside of full research papers. The main text and included appendices were analyzed, but no supplemental materials published outside of the main body (e.g., a linked website with additional findings) were examined.

Topically, these papers had to be about generative audio models. More specifically, a generative audio model had to be the primary focus of the paper. This meant that the paper either had to introduce a generative audio model/application, or analyze and discuss generative audio models as their focus of study. It was also important that these models were not conflated with another domain as their final output; for instance, though text-to-speech was included due to the output being in the audio domain, speech-to-text would be classified as a text generative model for the purposes of this paper and consequently excluded from analysis. Similarly, generative models resulting in video outputs were excluded due to conflating the visual domain with audio; the sole focus of the output of these models had to be entirely in the audio domain. Additionally, anything that was not generative in output (e.g., a classification model) was excluded.

Temporally, these papers had to be submitted or published in the last five years (at the time of research inception, this meant between February 1, 2018 and February 1, 2023). The reason for this was that this field is constantly evolving and any advances typically build upon the previous state-of-the-art performance which would rarely date prior to five years of research, especially prior to the introduction of the transformer in mid-2017 \cite{vaswani2017attention}. This means that research published over five years ago is not nearly as relevant to the field today as anything published recently.

\subsubsection{Keyword Search}

As a result of this aforementioned criteria (in Section \ref{sec:inclusion_criteria}), a keyword search was iteratively performed until the desired pool of research was included by the cast net. After many iterations of specific keywords such as ``music'', ``speech'', and ``sound'', the author eventually expanded the search to simply ``generative models'' and ``audio'' in order to comprehensively encompass as many potentially relevant articles as possible.\footnote{Initially, the search was targeted around ``deep generative audio models'', but it became clear quite quickly that the ``deep'' part of the term was too narrow and not widely adopted until recently (this search only resulted in 242 articles).} 

The search was initially focused on articles published in the Association for Computing Machinery (ACM) database, resulting in 444 potential articles meeting the criteria. After examining the articles included in this search, it quickly became clear that a large portion of the state-of-the-art research papers were either published outside of this domain at conferences such as Neural Information Processing Systems (NeurIPs) and the International Conference on Acoustics, Speech, \& Signal Processing (ICASSP), or simply not peer reviewed at an academic conference, yet widely respected in the field. Industry research is also growing in dominance in deep learning research \cite{ahmed2023growing}. It was important that these key papers from companies such as OpenAI and Google were included even though they may not be subject to peer review. Some of the most influential papers in this corpus were not peer reviewed yet still well regarded and well cited, like Jukebox \cite{dhariwal2020jukebox}$^\dag$ which was written by a team of researchers at OpenAI in 2020 that already had over 300 citations at the time of writing this paper. In order to include all of these essential papers in the corpus, the search was extended to include papers submitted to arXiv, which is an open-access non-peer reviewed archive for millions of research articles in the fields of physics, mathematics, computer science, quantitative biology, quantitative finance, statistics, electrical engineering and systems science, and economics \cite{arxiv_page}. This added an additional 440 articles to the initial screening pool, 65\% of which were peer reviewed. When screening these abstracts, it was noted whether the paper was peer reviewed or submitted to a conference/journal, however there were no notable trends among this specific dimension of whether the paper was peer reviewed.

The final query terms for ACM and arXiv were as follows:

\begin{itemize}
    \item \textbf{ACM}: [[[All: ``generative model''] OR [All: ``generative models''] OR [All: ``model generating'']] AND 
[All: ``audio'']]
AND [E-Publication Date: Past 5 years]''
    \item \textbf{arXiv}: ``(``generative model'' OR ``model generating'') AND 
``audio'' 
 date\_range: from 2018-02-01 to 2023-02-01''
\end{itemize}

\subsection{Title and Abstract Screening}

\begin{figure}[h]
  \centering
  \includegraphics[width=\linewidth]{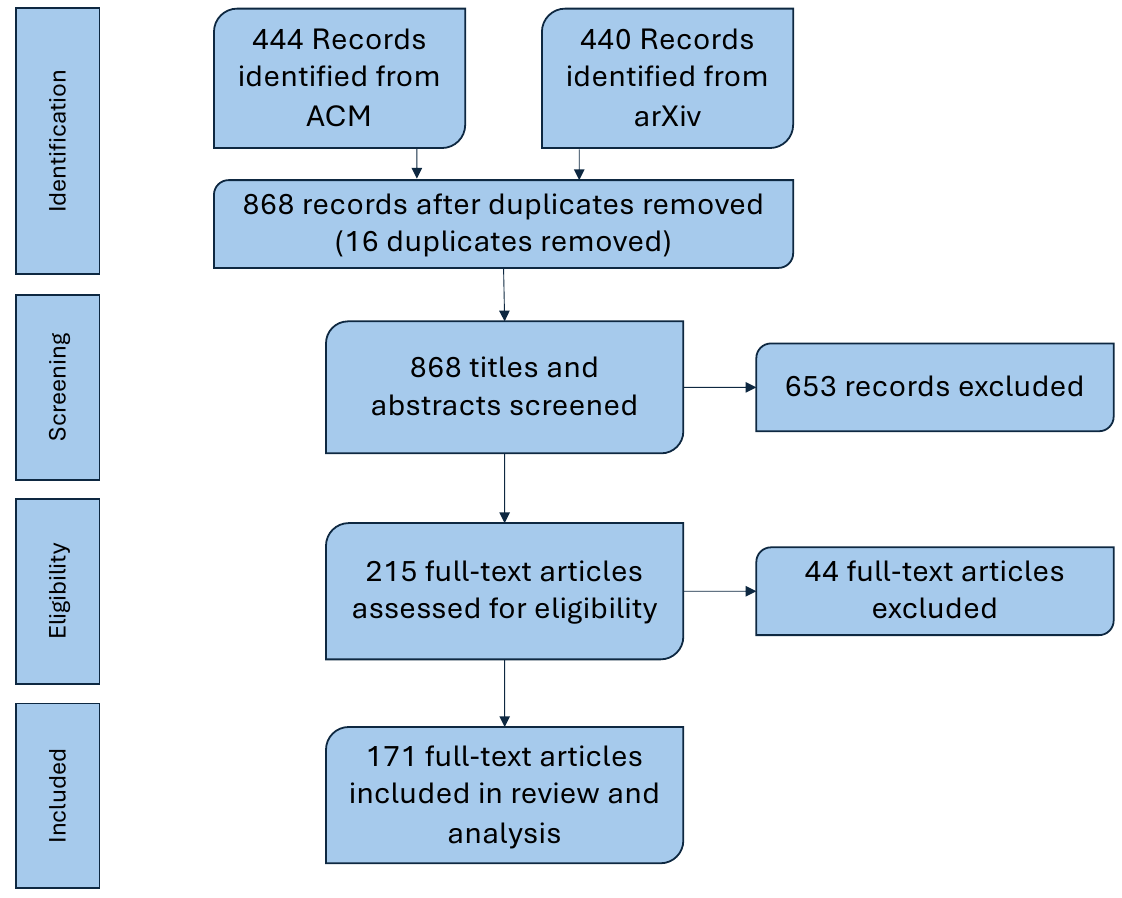}
  \caption{PRISMA Flow Diagram detailing the corpus screened and analyzed in the paper. After starting with an initial pool of 884 papers, 16 duplicates were removed, 653 records were excluded during the abstract screening stage, and 44 full-text articles were excluded during the eligibility review, resulting in 171 full-text articles for review and analysis.}
  \label{fig:prisma_flow}
  \Description{PRISMA Flow Diagram detailing the corpus screened and analyzed in the paper.}
\end{figure}

The author screened all 884 abstracts of the articles identified from the search (444 from ACM and 440 from arXiv). 16 duplicates were removed, and the remaining abstracts and titles were examined to determine eligibility. 653 articles were removed due to not meeting the criteria specified in Section \ref{sec:inclusion_criteria}. The vast majority of these articles were excluded due to not actually being about audio (46\%), but rather about text, vision, or some other central topic or output such as choreography. The second largest category of exclusion was due to papers not being a generative model (30\%) but rather something else such as a predictive or classification model. The remaining articles excluded at this stage were due to generating something in addition to audio like gestures or images (11\%), proposed a metric to evaluate the models or incrementally improve a specific aspect of the methodology without actually discussing any audio outputs or use cases (9\%), were all-encompassing and discussed generative models like text and vision in addition to audio (3\%), or were book chapters or abstracts instead of full papers (1\%). This is detailed in the screening section of Figure \ref{fig:prisma_flow}. The papers were excluded on a hierarchical list of criteria---for instance, it was possible that a paper was both not about audio and not a generative model, but would have been excluded for not meeting criteria (a) that the paper needed to be about audio, and thus these percentages for the later exclusion criteria are lower bounds. 

\subsection{Full Text Screening}

The title and abstract screening was quite thorough, so of the remaining 215 full-text articles only 44 additional papers were excluded resulting in 171 total papers for the analysis. Of these 44 papers, 20 (45\%) were removed for proposing a metric to evaluate the models or incrementally improve on a specific aspect of the methodology and not actually produce any audio output or apply to real data. 9 (20\%) were removed due to generating something in addition to audio such as gestures or video. 7 (16\%) were removed due to not being a generative model in nature but rather a classification or prediction model. 5 (11\%) were removed because they were not a full research paper but rather something like an extended abstract or book chapter. The remaining 3 (7\%) were excluded because their encompassing focus was too broad and included text and vision generative models in addition to audio models. Figure \ref{fig:prisma_flow} details the full flow diagram of papers from inception of the keyword search stage to final corpus.

The final corpus includes 140 papers from arXiv, 15 from the ACM database, and 16 which were in both databases. The vast majority of these papers ($n=122$; $71\%$) were peer reviewed at respected conferences and journals---only 18 papers from arXiv (13\% of arXiv papers) were not peer reviewed. The conferences/journals with at least five papers in the corpus were:

\begin{enumerate}
    \item Interspeech ($n=24$; $14\%$)
    \item Transactions on Audio, Speech, and Language Processing ($n=17$; $10\%$)
    \item International Conference on Acoustics, Speech and Signal Processing ($n=12$; $7\%$)
    \item International Society of Music Information Retrieval Conference ($n=9$; $5\%$)
    \item International Conference on Multimedia ($n=7$; $4\%$)
    \item Neural Information Processing Systems Conference ($n=6$; $4\%$)
    \item International Conference on Machine Learning ($n=5$; $3\%$)
\end{enumerate}

All content and thematic analyses discussed in the remainder of the paper were done by qualitative coding by the author. Research papers in the generative audio domain tend to fall either into two categories: music and speech. In this corpus, 77 (45\%) papers were about music and 103 (60\%) were about speech. Some papers include both ($n = 13$;  8\%), either by being datatype agnostic or by focusing on generating a singing voice. 4 papers (2\%) were about non-speech, non-music sounds, such as auxiliary sound effects like birds chirping and dogs barking. The summary of these topics by year can be found in Table \ref{tab:descriptives}.

Of the 103 speech papers, 48 (47\%) were about pure speech generation, 32 (31\%) were about text-to-speech, 24 (23\%) speech enhancement/denoising, 8 (8\%) accent conversion or style transfer, and 3 (3\%) audio inpainting or gap filling. There were additional topics scattered across papers, and the aforementioned categories were not mutually exclusive (for example one paper was about audio inpainting for controllable TTS models \cite{tae2021editts}$^\dag$). Of the 96 papers for which the language was either explicitly specified in the paper or able to be found through researching the datasets ($n=7, 7\%$  were not obtainable through the author's research), the vast majority utilized at least one English dataset ($n = 78$, 81\%), with 69 papers (72\%) exclusively using English training data. Of the remaining papers, 10 (10\%) used Mandarin, 9 (9\%) used Japanese, 3 (3\%) used Korean, and any other language specified was used in at most 2 papers. Notably, if a paper ever used a language other than English, it was explicitly stated. When the only language used was English, it was often necessary to research the datasets mentioned in order to determine the language utilized in the model, indicating further bias towards English in these models.

Of the 77 papers about music generation, 62 (81\%) used only instruments or non-lyric vocals, 10 (13\%) used pure vocals and lyrics, and 5 (6\%) used lyrics, vocals, and instruments. Of the 15 papers generating lyrics, 8 (53\%) explicitly mentioned the language chosen. 5 (63\%) used Mandarin Chinese, 4 (50\%) used English, and 1 (13\%) used Japanese, with 2 of the papers using both English and Mandarin datasets. 13 (17\%) of these music generation papers were evaluations of existing music generation models, and the remainder ($n=64, 83\%$) proposed a new model. 8 (10\%) focused on the HCI component of the models, 8 (10\%) used audio inpainting or gap filling, 5 (6\%) generated a musical score in addition to audio, and 4 (5\%) used a style transfer technique. The author did not note any additional topical trends within the corpus.

\begin{table}[ht]
\begin{tabular}{crrrrr}\toprule
 \multicolumn{6}{c}{\textbf{All Papers: Descriptive Statistics by Year }} \\
 \cmidrule(lr){1-6}
 \multicolumn{1}{c}{\textbf{Year}}&
 \multicolumn{1}{c}{\textbf{All Papers}}&
 \multicolumn{1}{c}{\textbf{Music}}&
 \multicolumn{1}{c}{\textbf{Speech}}&
 \multicolumn{1}{c}{\textbf{Both}}&
 \multicolumn{1}{c}{\textbf{Other}}\cr
 \midrule
 \textbf{2018} & \textbf{15} & 5 & 9 & 0 & 1\\
 \textbf{2019} & \textbf{34} & 17 & 18 & 2 & 1\\
 \textbf{2020} & \textbf{35} & 18 & 19 & 2 & 0\\
 \textbf{2021} & \textbf{41} & 19 & 27 & 6 & 1\\
 \textbf{2022} & \textbf{42} & 16 & 28 & 3 & 1\\
 \textbf{2023*} & \textbf{4} & 2 & 2 & 0 & 0\\
 \midrule
  \textbf{Sum} & \textbf{171} & 77 & 103 & 13 & 4\\ 
 \toprule
 \multicolumn{6}{c}{\textbf{Papers Discussing Negative Broader Impact}} \\
  \cmidrule(lr){1-6}
 \multicolumn{1}{c}{\textbf{Year}}&
 \multicolumn{1}{c}{\textbf{All Papers}}&
 \multicolumn{1}{c}{\textbf{Music}}&
 \multicolumn{1}{c}{\textbf{Speech}}&
 \multicolumn{1}{c}{\textbf{Both}}&
 \multicolumn{1}{c}{\textbf{Other}}\cr
 \midrule
 \textbf{2018} & \textbf{0} & 0 & 0 & 0 & 0\\
 \textbf{2019} & \textbf{1} & 0 & 1 & 0 & 0\\
 \textbf{2020} & \textbf{5} & 3 & 2 & 0 & 0\\
 \textbf{2021} & \textbf{4} & 2 & 3 & 1 & 0\\
 \textbf{2022} & \textbf{4} & 1 & 3 & 0 & 0\\
 \textbf{2023*} & \textbf{2} & 2 & 1 & 1 & 0\\
 \midrule
  \textbf{Sum} & \textbf{16} & 8 & 10 & 2 & 0\\ 
\bottomrule
\end{tabular}
\caption{\label{tab:descriptives}Descriptive statistics of research papers analyzed by year. Includes papers in music, speech, both music and speech, and other (non-music, non-speech) sounds. Note that 2023 only includes data for January.}
\vspace{-30pt}
\end{table}

\section{Analysis and Results}

\subsection{Overview}

Of these 171 papers, only 16 (9\%) discuss a negative broader impact, all of which were qualitatively coded by the author. 50\% ($n=8$) of these were in music papers, 63\% ($n=10$) were in speech papers, 13\% ($n=2$) were in both music and speech papers, and none were in the papers dealing with other non-music, non-speech audio. Temporally, there has been a slight increase over time of papers discussing negative broader impact, with none in 2018, 1 in 2019, and 4-5 each in 2020-2022. There have already been 2 papers in 2023 (with only one month of data for January) that discussed negative broader impacts, one of which was solely a music generation paper and the other discussed both music and speech generation. A full description of every paper discussing negative broader impacts can be found in Table \ref{tab:neg_impact_full}. 65\% ($n=112$) of the papers included in the corpus considered at least one positive broader impact of their work, so these researchers were considering broader impact---just not negative impact.

Even though only 9\% ($n = 16$) of the entire corpus discusses negative broader impacts, the papers that shine light on these ethical concerns do so in a manner that is inclusive to the vast majority of models in the domain. For instance, there are 32 text-to-speech (TTS) papers in the corpus, and though only 2 mention negative broader impacts, they do so in a manner that implicates all TTS models: Jaehyeon Kim et al. noted, ``TTS models could also be abused through cyber crimes such as fake news or phishing. It means that TTS models could be used to impersonate voices of celebrities for manipulating behaviours of people, or to imitate voices of someone’s friends or family for fraudulent purpose'' \cite{kim2020glow}$^\dag$. The negative broader impacts are not specific to the model proposed in the given paper, but rather to the entire domain of TTS and other speech models. Thus, the findings in this paper are not that 91\% of papers in the corpus have no need to discuss broader ethical impacts, but are far more likely to have neglected that discussion.

With the exception of one paper that was devoted almost entirely to a negative impact of audio models (energy consumption) \cite{douwes2021energy}$^\dag$, these papers tended to devote only 1-3 sentences to the potential negative impacts. Of these 16 papers, only 6 (38\%) papers included a short section devoted to ethical considerations or broader impact, and 2 of those were on the last page of the appendix. Of the remaining 9 papers, 2 (13\% of 16 papers) had 3 sentences devoted to negative impact, 5 (31\%) had 2 sentences, and 2 (13\%) had only part of one sentence. These sentences were primarily in the introduction ($n = 6$ papers; 38\%), discussion ($n = 4$; 25\%), and one sentence was in the conclusion (6\%).  

Jarringly, 2 of the 16 papers that discussed negative impacts explicitly mentioned that they did not have any intention to release their models or code due to the potential for misuse. Agostinelli et al., who created a text-to-music generation tool, stated: ``We acknowledge the risk of potential misappropriation of creative content associated to the use-case...We strongly emphasize the need for more future work in tackling these risks associated to music generation --- we have no plans to release models at this point'' \cite{agostinelli2023musiclm}$^\dag$, and Sungwon Kim et al., who created a TTS model, noted ``Given this potential misuse, we’ve decided not to release our code. Although we do not release the code, due to the adaptation ability of the diffusion-based model, we expect that the adaptive TTS technology is highly likely to be misused like Deepfake'' \cite{kim2022guided}$^\dag$. Two highly impactful papers both raise the alarm bells on the potential for misuse of generative audio models, and decided the safest mechanism for prevention of misuse was to not make their models or code available for public use. 

\begin{table*}[t]
\begin{tabular}{lclll}\toprule
 \multicolumn{5}{c}{\textbf{All Papers that Discuss Negative Broader Impact}} \\
 \cmidrule(lr){1-5}
  \multicolumn{5}{c}{\textbf{Music}} \\
 \cmidrule(lr){1-5}
  \multicolumn{1}{c}{\textbf{Reference}}&
 \multicolumn{1}{c}{\textbf{Year}}&
  \multicolumn{1}{c}{\textbf{Conf./Journal}}&
 \multicolumn{1}{c}{\textbf{Paper Topic}}&
 \multicolumn{1}{c}{\textbf{Negative Broader Impact}}\cr
 \midrule
 Frid et al. \cite{frid2020music}$^\dag$ & 2020 & CHI & Interface to generate music for videos & Loss of agency/authorship\\
 Zhao et al. \cite{zhao2020vertical}$^\dag$ & 2020 & N/A & Musical chord generation & Creativity stifling; Western bias\\
 Huang et al. \cite{huang2020ai}$^\dag$ & 2020 & ISMIR & Human+AI collaboration of & Creativity stifling;\\
 & & & music creation & Loss of agency/authorship\\
 Suh et al. \cite{Suh2021AI}$^\dag$ & 2021 & CHI & Human+AI collaboration of & Creativity stifling\\
  & & & music creation & \\
 Esling et al. \cite{esling2022challenges}$^\dag$ & 2022 & AIMCC & Music generation novelty & Copyright infringement;\\
 &&&&Creativity stifling\\
 Agostinelli et al. \cite{agostinelli2023musiclm}$^\dag$ & 2023 & N/A & Text-to-music generation & Copyright infringement; \\
 &&&&Cultural appropriation\\
 \cmidrule(lr){1-5}
 \multicolumn{5}{c}{\textbf{Speech}} \\
 \cmidrule(lr){1-5}
  \multicolumn{1}{c}{\textbf{Reference}}&
 \multicolumn{1}{c}{\textbf{Year}}&
  \multicolumn{1}{c}{\textbf{Conf./Journal}}&
 \multicolumn{1}{c}{\textbf{Paper Topic}}&
 \multicolumn{1}{c}{\textbf{Negative Broader Impact}}\cr
 \midrule
 Habib et al. \cite{habib2019semi}$^\dag$ & 2019 & ICLR & Text-to-speech & Fraud/Phishing; Misinformation\\
 Wang et al. \cite{Wang2020DeepSonar}$^\dag$ & 2020 & MM & Deep fakes for voices & Deepfakes; Fraud/Phishing; \\
 &&&&Security/Privacy\\
 Jaehyeon Kim et al. \cite{kim2020glow}$^\dag$ & 2020 & NeurIPS & Text-to-speech &  Deepfakes; Fraud/Phishing;\\\
 &&&&  Overuse of speaker data\\
 Li et al. \cite{Li2021Robust}$^\dag$ & 2021 & CCS & Adversarial attacks & Security/Privacy\\
 Sisman et al. \cite{Sisman2020overview}$^\dag$ & 2021 & TASLP & Voice conversion & Fraud/ Phishing\\
 Deng et al. \cite{deng2022v}$^\dag$ & 2022 & USENIXSS & Voice anonymization & Use of biometric data to identify people\\
 Sungwon Kim et al. \cite{kim2022guided}$^\dag$ & 2022 & N/A & Targeted user speech generation & Deepfakes; Fraud/Phishing;\\
 &&&&Security/Privacy\\
 Cho et al. \cite{cho2022attributable}$^\dag$ & 2022 & ICASSP & Model attribution & Fraud/Phishing; Security/Privacy\\
 \cmidrule(lr){1-5}
  \multicolumn{5}{c}{\textbf{Both Music and Speech}} \\
 \cmidrule(lr){1-5}
  \multicolumn{1}{c}{\textbf{Reference}}&
 \multicolumn{1}{c}{\textbf{Year}}&
  \multicolumn{1}{c}{\textbf{Conf./Journal}}&
 \multicolumn{1}{c}{\textbf{Paper Topic}}&
 \multicolumn{1}{c}{\textbf{Negative Broader Impact}}\cr
 \midrule
 Douwes et al. \cite{douwes2021energy}$^\dag$ & 2021 & ICASSP & Energy consumption of audio models & Energy consumption/Climate change\\
Huang et al. \cite{huang2023make}$^\dag$ & 2023 & N/A & Text-to-audio generation & Misinformation; Overuse of speaker data;\\ 
& & & & Unemployment\\
 \cmidrule(lr){1-5}
\bottomrule
\end{tabular}
\caption{\label{tab:neg_impact_full}Table describing the 16 papers in the corpus that discussed negative broader impacts, split by generative models concerning music, speech, and both music and speech. Table details the reference, year paper was published/submitted, conference or journal submitted to (or N/A if not peer reviewed), paper topic, and the negative broader impacts, organized topically by year.}
\end{table*}

\subsection{Negative Broader Impacts in Music}

Eight of the total papers in the corpus discussed potential negative broader impacts in the context of generative audio models for music. The main themes discussed in these were a loss of agency/authorship when creating the music, a general quelling of creativity, Western bias on the creation of music, copyright infringement, and cultural appropriation. They are discussed in more detail below, ordered chronologically by when papers first discussed the issue.

\vspace{-3pt}
\subsubsection{Loss of Agency and Authorship}

Two papers \cite{frid2020music, huang2020ai}$^\dag$ brought up the potential loss of agency that human creators would feel when creating music with the assistance of an AI generative model. Both of these papers were evaluations of existing models rather than creating their own generative model, and instead looked at the human-computer interaction (HCI) component of music generation models. Frid et al. noted that the co-creation of music from humans and machines ``raises interesting questions about autonomy, agency and authorship in human-AI interaction in creative practice'' and found that the human creators were hesitant to give the generative models too much control \cite{frid2020music}$^\dag$, indicating that musicians and creatives at large are wary of the recent focus on autonomous music generation. Huang et al. echoed this finding and found that novice musicians as well found it challenging to create jointly with AI and that ``users desire greater agency, control, and sense of authorship vis-a-vis the AI during co-creation'' \cite{huang2020ai}$^\dag$.

\vspace{-3pt}
\subsubsection{Creativity Stifling}

The most common potential negative impact discussed in the corpus was the stifling of creativity as a result of AI music generation \cite{esling2022challenges, huang2020ai, Suh2021AI, zhao2020vertical}$^\dag$. This focused on the repetitive nature of the music generation and that by limiting the creative output to possibilities of the model may result in a similar bound on human creativity. Suh et al. noted that these models ``may limit creative scope of humans'' \cite{Suh2021AI}$^\dag$, and Zhao et al. found that people ``may be not satisfied if the generated musical content tends to mimic the training set without exhibiting true creativity'' \cite{zhao2020vertical}$^\dag$. Both Huang et al. and Esling et al. suggested that a shift needs to be made toward steerable and interpretable models \cite{zhao2020vertical}$^\dag$, but ``introducing the notion of creativity in machine learning is difficult, as explicitly designing losses for creativity is an uphill battle'' \cite{esling2022challenges}$^\dag$. Many of the papers in the corpus position the generative audio models as a tool for assisting in the creativity process, so acknowledging the counterpoint is important as well.

\vspace{-3pt}
\subsubsection{Predominance of Western Bias}

Zhao et al. proposed a lightweight music generation model to generate instrumental music. In analyzing their output they found that their model was sensitive to Western music theory in that it ``it maintains the configuration of the circle of fifths; distinguishes major and minor keys from interval vectors, and manifests meaningful structures between music phases'' \cite{zhao2020vertical}$^\dag$. Machine learning models often perpetuate biases in the training data, and generative models are no different. It is important to be aware of the composition of the training data to understand what biases could be perpetuated.

\vspace{-3pt}
\subsubsection{Copyright Infringement}

Perhaps one of the most important considerations of generative music models---both ethically and potentially legally---was only discussed by two papers in the entire corpus: copyright infringement. There are many  legality questions surrounding the copyright of AI generated content. At least three lawsuits in early 2023 are currently discussing whether models trained on publicly available works constitute copyright infringement \cite{keller_2023}. Research in the text and vision domain is even geared toward specifically identifying to what degree generative models are memorizing training data \cite{carlini2022quantifying, mccoy2021much, somepalli2022diffusion} or are producing outputs with ``substantial similarity'' to items in the training set \cite{vyas2023provable}. However, in this corpus of generative audio models, only two papers discussed the potential for copyright infringement \cite{agostinelli2023musiclm, esling2022challenges}$^\dag$. Esling et al. focused their research on maximizing novelty in the music generation system in order to subvert the potential for copyright issues and increase creativity in their generation \cite{esling2022challenges}$^\dag$. Agostinelli et al. ``conducted a thorough study of memorization, adapting and extending a methodology used in the context of text-based LLMs'' in order to determine the degree to which their model memorized the training dataset and understand the potential for copyright infringement \cite{agostinelli2023musiclm}$^\dag$. Of the remaining 75 papers discussing generative music models (97\%), not one discussed the potential for copyright infringement or training data memorization.

\vspace{-3pt}
\subsubsection{Cultural Appropriation}

Generative audio models sometimes train on incomprehensible amounts of training data, and it follows that some of this training data comes from cultures outside the creator of the algorithm or users of the model. The ethical implications of this have been discussed in terms of computer vision; generative models make it easier to use content from marginalized cultures without any accompanying investment in or engagement from the community, even if the creators or users of the model are unaware of the use of that content \cite{rostamzadeh2021ethics}. Agostinelli et al. acknowledged that this extends to audio; ``The generated samples will reflect the biases present in the training data, raising the question about appropriateness for music generation for cultures underrepresented in the training data, while at the same time also raising concerns about cultural appropriation'' \cite{agostinelli2023musiclm}$^\dag$. A fundamental lack of understanding of model attribution will result in cultural appropriation if the training data contains content from marginalized communities.

\vspace{-3pt}
\subsection{Negative Broader Impacts in Speech}

Ten of the 16 papers in the corpus that discussed potential negative impacts did so in the context of speech generation. The ethical issues discussed exclusively relative to speech generation were fraud and phishing, misinformation and deepfake generation, security and privacy concerns, and the use of voice biometric data to identify people. They are discussed in more detail below, ordered chronologically by when papers first discussed the topic.

\subsubsection{Phishing and Fraud}

Six papers in the corpus discussed the potential misuse of the generative speech models for committing phishing and fraud. Habib et al. noted that ``progress in controllability raises the prospect that bad actors may misuse the technology either for misinformation or to commit fraud'' \cite{habib2019semi}$^\dag$. Wang et al. echoed this concern by noting that these bad actors could use victims' voices for fraudulent purposes \cite{Wang2020DeepSonar}$^\dag$, and Sisman et al. specifically called out the need for anti-spoofing countermeasures as ``voice conversion technology could be misused for attacking speaker verification systems'' \cite{Sisman2020overview}$^\dag$. Fraud can occur whenever an audio model targets the speech of an individual and is able to impersonate them, either for formal voice verification fraud or impersonating people close to the victim in order to mislead them. Text-to-speech models especially have the potential to be misused by bad actors due to the ease of guiding model output with a target speaker as the medium; Sungwon Kim et al. acknowledged that their model, Guided-TTS 2, was ``likely to be misused as voice phishing for individuals'' \cite{kim2022guided}$^\dag$ and chose not to release their code or models to the public. Cho et al. proposed a model that was designed to focus on attribution, which in their words is ``much more difficult to spoof'' \cite{cho2022attributable}$^\dag$ compared to non-attributable models.

\subsubsection{Misinformation and Deepfakes} \label{sec:misinformation_deepfakes}

A slightly nuanced aspect of speech generative models' ability to impersonate victims exists when the victims are famous and the model misuse can take the form of misinformation or deepfakes. As Wang et al. noted ``some attackers and criminals misuse them for illegal purposes like a politician giving an unreal statement, which may cause a regional crisis'' \cite{Wang2020DeepSonar}$^\dag$. Jaehyeon Kim et al. noted that TTS models were particularly vulnerable to deepfakes, stating that ``because of the ability to synthesize natural speech, the TTS models...could be used to impersonate voices of celebrities for manipulating behaviours of people'' \cite{kim2020glow}$^\dag$. This risk is amplified when the needed length of speech to train a targeted speaker output is small---Sungwon Kim et al. remarked that a ``10-second untranscribed speech for the target speaker is easy to obtain through recording or YouTube clips for celebrities, and the contribution of [their model] that reduces the data required for high-quality adaptive TTS makes a lot of room for misuse'' \cite{kim2022guided}$^\dag$. As these models continue to become easier to use, the prevalence of deepfakes and misinformation online will continue to grow.

\vspace{-3mm}
\subsubsection{Security and Privacy}

Three papers in the corpus discussed the potential for risk to security and privacy of individuals as a result of speech generative models, especially when they only require small segments of training data to produce a realistic voice of a targeted speaker. Cho et al. stated, ``these models and their synthetic contents inevitably pose a variety of threats regarding privacy'' \cite{cho2022attributable}$^\dag$, and Wang et al. noted that the ease of use of these models results in ``security and privacy concerns to everyone while we are enjoying the fun of these synthesized fakes'' \cite{Wang2020DeepSonar}$^\dag$. Sungwon Kim et al. asserted that due to the short length of speech necessary to target a speaker, this type of content can be easily obtained and resultingly ``have a fatal effect on the security system through voice'' \cite{kim2022guided}$^\dag$. In addition to targeted impersonation attacks, there are also machine-induced audio attacks on intelligent audio systems such as hidden voice commands; Li et al. designed a solution to detect targeted machine-induced audio attacks in order to add some level of security to audio-triggered devices and mechanisms \cite{Li2021Robust}$^\dag$.

\subsubsection{Non-consensual Use of Biometric Data}

Voiceprint is a type of audio finger-printing that has been around for decades that can identify individuals with varying levels of accuracy \cite{delac2004survey, kersta1962voiceprint}. Though there have been recent efforts to protect biometric data such as the European General Data Protection Regulation (GDPR), an immense amount of voice data is accumulated daily on social media apps like TikTok and Facebook \cite{deng2022v}$^\dag$. Deng. et al. designed a model to protect voiceprint through the anonymization of voice data. They acknowledged that this could be abused, and stated that they would take proper measures to prevent the abuse of the anonymization system \cite{deng2022v}$^\dag$.

\subsection{Negative Broader Impacts in Both Music and Speech}

Finally, there were two papers that discussed negative broader impacts of generative models  both in terms of music and speech generation from an output-agnostic standpoint. In these papers, three main topics were presented that were neither specific to speech nor music models. These topics were the energy consumption of audio models, overuse of speaker data, and unemployment. These are discussed below. One of these papers discussed misinformation, but only in context of speech models so it is discussed above in Section \ref{sec:misinformation_deepfakes}.

\subsubsection{Energy Consumption of Generative Audio Models}

There was one paper in the entire corpus (0.6\%) that discussed the carbon footprint of audio models, and the entire paper was dedicated to the topic \cite{douwes2021energy}$^\dag$. Douwes et al. proposed a new multi-objective measure to evaluate deep generative audio models that takes into account both the quality and energy consumption of the model. 

There are two types of energy consumption of a generative model: the energy required to train and to generate samples. Current research points to machine learning models being at risk of becoming a significant contributor to climate change, and proposes the total energy consumption and carbon emissions of training these models be reported alongside the other standard suite of metrics \cite{anthony2020carbontracker}. This energy consumption also varies by region and country in which the electricity is generated---Anthony et al. find that a single training session of a standard medical image segmentation model trained in Estonia would emit about 61 times as much carbon dioxide equivalent on the basis of their global-warming potential versus a model trained in Sweden, or in laypersons's terms the difference between travelling 9.04 km by car versus 0.14 km by car \cite{anthony2020carbontracker}. In this corpus, Douwes et al. focused on specifically increasing awareness of the energy consumption of generative audio models and elevating computational complexity and carbon footprint in line with other model quality metrics \cite{douwes2021energy}$^\dag$. Though this is the only paper in the corpus that discusses the carbon footprint of generative audio models, this is a metric relevant for every single model. 

\vspace{-7pt}
\subsubsection{Overuse of Speaker Data}

There is a tendency across all various realms of machine learning and AI to reuse publicly available datasets, in fact 76\% of the papers in the corpus that used data to train models utilized datasets that were already available. Many of these datasets containing recordings of human voices are only comprised of a few human beings. Jaehyeon Kim et al. described this concern; ``Many corpus for speech synthesis contain speech data uttered by a handful of speakers. Without the detailed consideration and restriction about the range of uses the TTS models have, the voices of the speakers could be overused than they might expect'' \cite{kim2020glow}$^\dag$, and Huang et al. echoed this exact sentiment: ``the voices in the recordings might be overused than they expect'' \cite{huang2023make}$^\dag$. When signing up to record speech for a singular research project, people may not realize the potential extent to which their voices could be used in future models and other outputs.

\vspace{-7pt}
\subsubsection{Unemployment}

Finally, Huang et al. discussed unemployment as a potential result of lowering the barriers to entry for various audio generation jobs. They postulated that their model ``lowers the requirements for high-quality text-to-audio synthesis, which may cause unemployment for people with related occupations, such as sound engineers and radio hosts'' \cite{huang2023make}$^\dag$. There are varying findings on the macro-level effect of artificial intelligence on employment, which has found to depend on inflation and can be netural or positive for employment \cite{mutascu2021artificial}. However, on a micro-level different innovations of AI such as a generative music model can certainly displace current jobs as noted by Huang et al. \cite{huang2023make}$^\dag$. Though current research in economics suggest that AI could instead increase a demand for jobs in these domains, the types of jobs will shift as a result of the automation---called ``job displacement'' \cite{acemoglu2018artificial}---which is worth noting in papers proposing models that could displace current jobs.

\section{Discussion}

These findings highlight the necessity of generative audio researchers to place a greater emphasis on the consideration of the negative impacts of their work. The severity of the negative impacts highlighted by the few papers that acknowledge them indicates that the vast majority of these researchers of generative audio models are not considering negative impact due to negligence, rather than lack of necessity. An argument could be made that computer scientists are not obliged to think in terms of broader societal impacts, however, the vast majority of them are already doing so. The catch is that they are only thinking in terms of \textit{positive} societal impacts; 65\% ($n=112$) of the papers included in the corpus considered at least on positive broader impact of their work. These researchers are already inclined to consider broader impact; they just need to consider negative impact as well.

The author first acknowledges the limitations of the corpus. One limitation is that this research was focused on the generative audio domain in isolation---it did not include videos or any other multimedia audio synthesis. The potential landscape for negative impact in this multimedia is compounded, and things like realistic video deepfakes \cite{Mirsky2021Creation} can become potentially more harmful. Another limitation is inherent to the scoping of the SLR: the keyword search in both databases likely did not encompass every single generative audio paper published in the last five years. Though the author attempted to cast as wide a net as possible for the initial identification of articles, it is inevitable that some papers eluded the search and thus were not included in this analysis. 

Revisiting the areas of ethical and social risks of harm in language models (LMs) established by Weidinger et al. \cite{weidinger2021ethical} discussed in Section \ref{sec:broader_impact}, this systematic literature review uncovered harms in all of the categories established in this taxonomy. Discrimination, exclusion, and toxicity harms can include cultural appropriation and the predominance of Western bias found in this review, however this area of harms can extend much further than what was found in the discussions in the corpus. Weidinger et al.'s classification of information hazards directly translated to security and privacy concerns of audio models. Misinformation harms were also able to be extended from text to audio, specifically in speech models. Malicious uses can take different forms in audio models than they do in LMs, such as deepfakes, but the concern of fraud and phishing can be examined in a similar manner as that of LMs. Human-computer interaction harms varied slightly due to the focus on the loss of agency and authorship and creativity, whereas LMs focused on unsafe use due to users misjudging or mistakenly trusting the model. Automation, access, and environmental harms encompassed the energy consumption of audio models, unemployment, overuse of speaker data, use of biometric data to identify people, and even copyright infringement in the sense that it undermines creative economies. 

The 16 papers that mentioned potential negative impacts brought to light a wide variety of ethical implications that the field at large needs to consider going forward during the design process, the implementation of their models, and the publication and \textit{publicization} of their research. Two papers, one in music generation \cite{agostinelli2023musiclm}$^\dag$ and one in speech \cite{kim2022guided}$^\dag$, decided that the potential risk of misuse by bad actors was too great to release their models to the public. This is a consideration that every researcher working on deep generative audio models should make prior to allowing their models to be public facing. If the potential risks outweigh the benefits, then it may not be justifiable to release code or models.

At a minimum, researchers focusing on generative audio models going forward need to consider the set of impacts discussed in this paper. For research generating music, that means loss of agency and authorship of the human creator, stifling of creativity, a predominance of Western bias in their data and any other data biases for that matter, the possibility of copyright infringement, and cultural appropriation. For speech models, it is essential to consider misuse pertaining to phishing and fraud, misinformation and deepfakes, security and privacy concerns of these models, and non-consensual use of biometric data. All generative audio models need to be aware of their carbon footprint and potential energy consumption---ideally explicitly listing these metrics in tandem with other representations of quality. They also need to consider the overuse of speaker (and singer/musician) data being used in much larger corpora and models beyond the immediate use-case of the model, and the potential job displacement of people who are currently employed to perform the task that the model could be replacing.

This is not meant to be an exhaustive list of potential impacts---merely a minimum set of considerations for generative audio models going forward. It should be seen as a starting point to begin thinking in terms of broader impact beyond simply the potential benefit to society. The potential impact on society should be considered at all stages of the research process, and researchers need to take steps to prevent potential harm. This paper does not simply call for more researchers to put a disclaimer at the end of their research papers, though that is a necessary aspect as well. Generative audio researchers need to consider potential negative impact all throughout their research and ensure that all stages of their work---from brainstorming to implementation and publication---are conducted with care and consideration for society at large. 

\section{Conclusion}

In this paper, the author conducted a systematic literature review of research papers in the generative audio domain in order to understand both the degree to which current researchers consider the negative broader impact of their work and also thematically evaluate the types of ethical implications discussed. The findings indicate that less than 10\% of research papers discuss any negative broader impact in their work, even though 65\% consider potential positive broader impacts. This small percentage is not reflective of the degree of necessity of considering negative impact because the issues brought to light by the few papers doing so are raising serious ethical implications and concerns like the potential for fraud, deepfakes, and copyright infringement. Two of the papers even explicitly noted they had no plans to release their models or code due to the strong potential for misuse. This paper quantifies the lack of ethical consideration of researchers in the generative audio domain at a critical point in time and lays the groundwork for future work in the field to consider potential negative impacts as work in this field progresses.

\begin{acks}
The author would like to thank Michelle Shumate, Nick Diakopoulos, and Mackenzie Jorgensen for their helpful feedback.
\end{acks}

\bibliographystyle{ACM-Reference-Format}
\bibliography{references}

\appendix

\section{References for Works Included in Systematic Literature Review}

\vspace{2pt}
This appendix includes the full citation for each of the 171 works included in the full text analysis in the systematic literature review. If a paper was additionally cited in text in the main body of the paper with a dagger$^\dag$ symbol, that citation will align with the standard references above and their full citation is listed again below alphabetically by last name with a numbering system that does not align with in-text citations.
\begin{enumerate}[ {[}1{]} ]
\item Andrea Agostinelli et al. “MusicLM: Generating Music From Text”. In: arXiv preprint arXiv:2301.11325 (2023). 
\item Cyran Aouameur, Philippe Esling, and Ga\"etan Hadjeres. “Neural drum ma- chine: An interactive system for real-time synthesis of drum sounds”. In: arXiv preprint arXiv:1907.02637 (2019). 
\item Sercan Arik et al. “Neural voice cloning with a few samples”. In: Advances in neural information processing systems 31 (2018). 
\item Yoshiaki Bando et al. “Statistical speech enhancement based on probabilistic integration of variational autoencoder and non-negative matrix factorization”. In: 2018 IEEE International Conference on Acoustics, Speech and Signal Processing (ICASSP). IEEE. 2018, pp. 716–720. 
\item Th\'eis Bazin and Ga\"etan Hadjeres. “Nonoto: A model-agnostic web interface for interactive music composition by inpainting”. In: arXiv preprint arXiv:1907.10380 (2019). 
\item Xiaoyu Bie et al. “A benchmark of dynamical variational autoencoders applied to speech spectrogram modeling”. In: arXiv preprint arXiv:2106.06500 (2021). 
\item Xiaoyu Bie et al. “Unsupervised speech enhancement using dynamical varia- tional autoencoders”. In: IEEE/ACM Transactions on Audio, Speech, and Lan- guage Processing 30 (2022), pp. 2993–3007. 
\item Miko\l aj Bi\'nkowski et al. “High fidelity speech synthesis with adversarial networks”. In: arXiv preprint arXiv:1909.11646 (2019). 
\item Adrien Bitton, Philippe Esling, and Axel Chemla-Romeu-Santos. “Modulated variational auto-encoders for many-to-many musical timbre transfer”. In: arXiv preprint arXiv:1810.00222 (2018). 
\item Adrien Bitton et al. “Assisted sound sample generation with musical con- ditioning in adversarial auto-encoders”. In: arXiv preprint arXiv:1904.06215 (2019). 
\item Bajibabu Bollepalli, Lauri Juvela, and Paavo Alku. “Generative adversarial network-based glottal waveform model for statistical parametric speech syn- thesis”. In: arXiv preprint arXiv:1903.05955 (2019).
\item Gilles Boulianne. “A study of inductive biases for unsupervised speech repre- sentation learning”. In: IEEE/ACM Transactions on Audio, Speech, and Language Processing 28 (2020), pp. 2781–2795.
\item Korneel van den Broek. “Mp3net: coherent, minute-long music generation from raw audio with a simple convolutional GAN”. In: arXiv preprint arXiv:2101.04785 (2021).
\item Gino Brunner et al. “Symbolic music genre transfer with cyclegan”. In: 2018 ieee 30th international conference on tools with artificial intelligence (ictai). IEEE. 2018, pp. 786–793.
\item Zexin Cai, Yaogen Yang, and Ming Li. “Cross-lingual multispeaker text-to- speech under limited-data scenario”. In: arXiv preprint arXiv:2005.10441 (2020).
\item Antoine Caillon and Philippe Esling. RAVE: A variational autoencoder for fast and high-quality neural audio synthesis. Dec. 2021.
\item Pablo Samuel Castro. “Performing structured improvisations with pre-trained deep learning models”. In: arXiv preprint arXiv:1904.13285 (2019).
\item Pritish Chandna et al. “LoopNet: Musical loop synthesis conditioned on intu- itive musical parameters”. In: ICASSP 2021-2021 IEEE International Conference on Acoustics, Speech and Signal Processing (ICASSP). IEEE. 2021, pp. 3395–3399.
\item Gong Chen et al. “Musicality-novelty generative adversarial nets for algorithmic composition”. In: Proceedings of the 26th ACM international conference on Multimedia. 2018, pp. 1607–1615.
\item Mingjie Chen and Thomas Hain. “Unsupervised acoustic unit representation learning for voice conversion using wavenet auto-encoders”. In: arXiv preprint arXiv:2008.06892 (2020).
\item Nanxin Chen et al. “Wavegrad 2: Iterative refinement for text-to-speech synthesis”. In: arXiv preprint arXiv:2106.09660 (2021).
\item Hyunjae Cho et al. “SANE-TTS: Stable And Natural End-to-End Multilingual Text-to-Speech”. In: arXiv preprint arXiv:2206.12132 (2022).
\item Yongbaek Cho et al. “Attributable Watermarking of Speech Generative Models”. In: ICASSP 2022-2022 IEEE International Conference on Acoustics, Speech and Signal Processing (ICASSP). IEEE. 2022, pp. 3069–3073.
\item Jiangyi Deng et al. “V-Cloak: Intelligibility-, Naturalness-\& Timbre-Preserving Real-Time Voice Anonymization”. In: arXiv preprint arXiv:2210.15140 (2022).
\item Prafulla Dhariwal et al. “Jukebox: A generative model for music”. In: arXiv preprint arXiv:2005.00341 (2020).
\item Sander Dieleman, Aaron van den Oord, and Karen Simonyan. “The challenge of realistic music generation: modelling raw audio at scale”. In: Advances in Neural Information Processing Systems 31 (2018).
\item Constance Douwes, Philippe Esling, and Jean-Pierre Briot. “Energy Consumption of Deep Generative Audio Models”. In: arXiv preprint arXiv:2107.02621 (2021).
\item Zhihao Du, Xueliang Zhang, and Jiqing Han. “A joint framework of denoising autoencoder and generative vocoder for monaural speech enhancement”. In: IEEE/ACM Transactions on Audio, Speech, and Language Processing 28 (2020), pp. 1493–1505.
\item Jesse Engel et al. “DDSP: Differentiable digital signal processing”. In: arXiv preprint arXiv:2001.04643 (2020).
\item Philippe Esling et al. “Challenges in creative generative models for music: a divergence maximization perspective”. In: arXiv preprint arXiv:2211.08856 (2022).
\item Cundi Fang, Zhiyong Li, and Zhihao Ye. “Automatic Music Creation Based on Bayesian Networks”. In: Proceedings of the 2020 4th International Conference on Vision, Image and Signal Processing. 2020, pp. 1–6.
\item Lucas Fenaux and Maria Juliana Quintero. BumbleBee: A Transformer for Music. July 2021.
\item Emma Frid, Celso Gomes, and Zeyu Jin. “Music creation by example”. In: Proceedings of the 2020 CHI conference on human factors in computing systems. 2020, pp. 1–13.
\item Benjamin Genchel, Ashis Pati, and Alexander Lerch. “Explicitly conditioned melody generation: A case study with interdependent rnns”. In: arXiv preprint arXiv:1907.05208 (2019).
\item Jon Gillick et al. “Learning to groove with inverse sequence transformations”. In: International Conference on Machine Learning. PMLR. 2019, pp. 2269–2279.
\item Gal Greshler, Tamar Shaham, and Tomer Michaeli. “Catch-a-waveform: Learning to generate audio from a single short example”. In: Advances in Neural Information Processing Systems 34 (2021), pp. 20916–20928.
\item Yu Gu and Yongguo Kang. “Multi-task WaveNet: A multi-task generative model for statistical parametric speech synthesis without fundamental frequency conditions”. In: arXiv preprint arXiv:1806.08619 (2018).
\item Raza Habib et al. “Semi-supervised generative modeling for controllable speech synthesis”. In: arXiv preprint arXiv:1910.01709 (2019).
\item Ga\"etan Hadjeres and L\'eopold Crestel. The Piano Inpainting Application. July 2021
\item Sangjun Han et al. “Symbolic Music Loop Generation with Neural Discrete Representations”. In: arXiv preprint arXiv:2208.05605 (2022).
\item Zack Hodari, Oliver Watts, and Simon King. “Using generative modelling to produce varied intonation for speech synthesis”. In: 10th ISCA Workshop on Speech Synthesis (SSW 10). Sept. 2019, pp. 239–244.
\item Joanna Hong et al. “Speech reconstruction with reminiscent sound via visual voice memory”. In: IEEE/ACM Transactions on Audio, Speech, and Language Processing 29 (2021), pp. 3654–3667.
\item Yukiya Hono et al. “Hierarchical multi-grained generative model for expressive speech synthesis”. In: arXiv preprint arXiv:2009.08474 (2020).
\item Yukiya Hono et al. “PeriodNet: A non-autoregressive waveform generation model with a structure separating periodic and aperiodic components”. In: ICASSP 2021-2021 IEEE International Conference on Acoustics, Speech and Signal Processing (ICASSP). IEEE. 2021, pp. 6049–6053.
\item Yukiya Hono et al. “Sinsy: A deep neural network-based singing voice synthesis system”. In: IEEE/ACM Transactions on Audio, Speech, and Language Processing 29 (2021), pp. 2803–2815.
\item Po-chun Hsu and Hung-yi Lee. “WG-WaveNet: Real-time high-fidelity speech synthesis without GPU”. In: arXiv preprint arXiv:2005.07412 (2020).
\item Po-chun Hsu et al. “Parallel Synthesis for Autoregressive Speech Generation”. In: arXiv preprint arXiv:2204.11806 (2022).
\item Wei-Ning Hsu et al. “Hierarchical generative modeling for controllable speech synthesis”. In: arXiv preprint arXiv:1810.07217 (2018).
\item Cheng-Zhi Anna Huang et al. “AI song contest: Human-AI co-creation in songwriting”. In: arXiv preprint arXiv:2010.05388 (2020).
\item Renjie Huang et al. “Melody Generation with Emotion Constraint”. In: Pro- ceedings of the 2021 5th International Conference on Electronic Information Technology and Computer Engineering. 2021, pp. 1598–1603.
\item Rongjie Huang et al. “GenerSpeech: Towards Style Transfer for Generalizable Out-Of-Domain Text-to-Speech Synthesis”. In: arXiv preprint arXiv:2205.07211 (2022).
\item Rongjie Huang et al. “Make-An-Audio: Text-To-Audio Generation with Prompt- Enhanced Diffusion Models”. In: arXiv preprint arXiv:2301.12661 (2023).
\item Rongjie Huang et al. “Prodiff: Progressive fast diffusion model for high-quality text-to-speech”. In: Proceedings of the 30th ACM International Conference on Multimedia. 2022, pp. 2595–2605.
\item Rongjie Huang et al. “Singgan: Generative adversarial network for high-fidelity singing voice generation”. In: Proceedings of the 30th ACM International Con- ference on Multimedia. 2022, pp. 2525–2535.
\item Hsiao-Tzu Hung et al. “Improving automatic jazz melody generation by transfer learning techniques”. In: 2019 Asia-Pacific Signal and Information Processing Association Annual Summit and Conference (APSIPA ASC). IEEE. 2019, pp. 339– 346.
\item Mumin Jin et al. Voice-preserving Zero-shot Multiple Accent Conversion. Nov. 2022.
\item Lauri Juvela et al. “Speaker-independent raw waveform model for glottal excitation”. In: arXiv preprint arXiv:1804.09593 (2018).
\item Hirokazu Kameoka et al. “ConvS2S-VC: Fully convolutional sequence-to- sequence voice conversion”. In: IEEE/ACM Transactions on audio, speech, and language processing 28 (2020), pp. 1849–1863.
\item Hirokazu Kameoka et al. “Nonparallel voice conversion with augmented classi- fier star generative adversarial networks”. In: IEEE/ACM Transactions on Audio, Speech, and Language Processing 28 (2020), pp. 2982–2995.
\item Minki Kang, Dongchan Min, and Sung Ju Hwang. “Any-speaker Adaptive Text- To-Speech Synthesis with Diffusion Models”. In: arXiv preprint arXiv:2211.09383 (2022).
\item Anurag Katakkar and Alan W Black. “Towards Language Modelling in the Speech Domain Using Sub-word Linguistic Units”. In: arXiv preprint arXiv:2111.00610 (2021).
\item Navjot Kaur and Paige Tuttosi. “Time out of Mind: Generating Rate of Speech conditioned on emotion and speaker”. In: arXiv e-prints (2023), arXiv–2301.
\item Tom Kenter et al. “CHiVE: Varying prosody in speech synthesis with a lin- guistically driven dynamic hierarchical conditional variational network”. In: International Conference on Machine Learning. PMLR. 2019, pp. 3331–3340.
\item Jaehyeon Kim, Jungil Kong, and Juhee Son. “Conditional variational autoen- coder with adversarial learning for end-to-end text-to-speech”. In: International Conference on Machine Learning. PMLR. 2021, pp. 5530–5540.
\item Jaehyeon Kim et al. “Glow-TTS: A generative flow for text-to-speech via monotonic alignment search”. In: Advances in Neural Information Processing Systems 33 (2020), pp. 8067–8077.
\item Sungwon Kim, Heeseung Kim, and Sungroh Yoon. “Guided-TTS 2: A diffusion model for high-quality adaptive text-to-speech with untranscribed data”. In: arXiv preprint arXiv:2205.15370 (2022).
\item Eunjeong Stella Koh, Shlomo Dubnov, and Dustin Wright. “Rethinking re- current latent variable model for music composition”. In: 2018 IEEE 20th In- ternational Workshop on Multimedia Signal Processing (MMSP). IEEE. 2018, pp. 1–6. 
\item Jungil Kong, Jaehyeon Kim, and Jaekyoung Bae. “Hifi-gan: Generative adver- sarial networks for efficient and high fidelity speech synthesis”. In: Advances in Neural Information Processing Systems 33 (2020), pp. 17022–17033. 
\item Junghyun Koo, Seungryeol Paik, and Kyogu Lee. “End-to-end Music Remas- tering System Using Self-supervised and Adversarial Training”. In: ICASSP 2022-2022 IEEE International Conference on Acoustics, Speech and Signal Pro- cessing (ICASSP). IEEE. 2022, pp. 4608–4612. 
\item Daniel Korzekwa et al. “Interpretable deep learning model for the detection and reconstruction of dysarthric speech”. In: arXiv preprint arXiv:1907.04743 (2019).
\item Felix Kreuk et al. “Audiogen: Textually guided audio generation”. In: arXiv preprint arXiv:2209.15352 (2022). 
\item Ohsung Kwon et al. “Effective parameter estimation methods for an excitnet model in generative text-to-speech systems”. In: arXiv preprint arXiv:1905.08486 (2019).
\item Chae Young Lee et al. Conditional WaveGAN. Sept. 2018. 
\item Sang-gil Lee et al. BigVGAN: A Universal Neural Vocoder with Large-Scale Training. Feb. 2023.
\item Seokjin Lee et al. “Conditional variational autoencoder to improve neural audio synthesis for polyphonic music sound”. In: arXiv preprint arXiv:2211.08715 (2022). 
\item Jean-Marie Lemercier et al. “Analysing Diffusion-based Generative Approaches versus Discriminative Approaches for Speech Restoration”. In: arXiv preprint arXiv:2211.02397 (2022).
\item Yinghao Aaron Li, Cong Han, and Nima Mesgarani. “StyleTTS: A Style-Based Generative Model for Natural and Diverse Text-to-Speech Synthesis”. In: arXiv preprint arXiv:2205.15439 (2022). 
\item Zhuohang Li et al. “Robust Detection of Machine-Induced Audio Attacks in Intelligent Audio Systems with Microphone Array”. In: Proceedings of the 2021 ACM SIGSAC Conference on Computer and Communications Security. CCS ’21. Virtual Event, Republic of Korea: Association for Computing Machinery, 2021, 1884–1899. isbn: 9781450384544. doi: 10.1145/3460120.3484755. 
\item Xia Liang, Junmin Wu, and Jing Cao. “MIDI-Sandwich2: RNN-based Hierar- chical Multi-modal Fusion Generation VAE networks for multi-track symbolic music generation”. In: arXiv preprint arXiv:1909.03522 (2019).
\item Xia Liang, Junmin Wu, and Yan Yin. “MIDI-Sandwich: Multi-model Multi-task Hierarchical Conditional VAE-GAN networks for Symbolic Single-track Music Generation”. In: arXiv preprint arXiv:1907.01607 (2019). 
\item Jen-Yu Liu et al. “Score and lyrics-free singing voice generation”. In: arXiv preprint arXiv:1912.11747 (2020).
\item Jinglin Liu et al. “Diffsinger: Singing voice synthesis via shallow diffusion mechanism”. In: Proceedings of the AAAI Conference on Artificial Intelligence. Vol. 36. 10. 2022, pp. 11020–11028. 
\item Jinglin Liu et al. “Learning the Beauty in Songs: Neural Singing Voice Beauti- fier”. In: arXiv preprint arXiv:2202.13277 (2022).
\item Songxiang Liu, Dan Su, and Dong Yu. “Diffgan-TTS: High-fidelity and efficient text-to-speech with denoising diffusion gans”. In: arXiv preprint arXiv:2201.11972 (2022). 
\item Xubo Liu et al. “Conditional sound generation using neural discrete time- frequency representation learning”. In: 2021 IEEE 31st International Workshop on Machine Learning for Signal Processing (MLSP). IEEE. 2021, pp. 1–6.
\item Ryan Louie, Jesse Engel, and Anna Huang. “Expressive communication: A common framework for evaluating developments in generative models and steering interfaces”. In: arXiv preprint arXiv:2111.14951 (2021). 
\item Yen-Ju Lu, Yu Tsao, and Shinji Watanabe. “A study on speech enhancement based on diffusion probabilistic model”. In: 2021 Asia-Pacific Signal and Infor- mation Processing Association Annual Summit and Conference (APSIPA ASC). IEEE. 2021, pp. 659–666. 
\item Yen-Ju Lu et al. Conditional Diffusion Probabilistic Model for Speech Enhance- ment. Feb. 2022.
\item Jing Luo et al. “MG-VAE: Deep Chinese folk songs generation with specific regional styles”. In: Proceedings of the 7th Conference on Sound and Music Technology (CSMT) Revised Selected Papers. Springer. 2020, pp. 93–106. 
\item Ang Lv et al. “Re-creation of Creations: A New Paradigm for Lyric-to-Melody Generation”. In: arXiv e-prints (2022), arXiv–2208.
\item Andr\'es Marafioti et al. “A context encoder for audio inpainting”. In: IEEE/ACM Transactions on Audio, Speech, and Language Processing 27.12 (2019), pp. 2362– 2372. 
\item Ollie McCarthy and Zohaib Ahmed. “HooliGAN: Robust, high quality neural vocoding”. In: arXiv preprint arXiv:2008.02493 (2020). 
\item Dongchan Min et al. “Meta-stylespeech: Multi-speaker adaptive text-to-speech generation”. In: International Conference on Machine Learning. PMLR. 2021, pp. 7748–7759.
\item Huaiping Ming et al. “Feature reinforcement with word embedding and parsing information in neural TTS”. In: arXiv preprint arXiv:1901.00707 (2019).
\item Gautam Mittal et al. Symbolic Music Generation with Diffusion Models. Nov. 2021. 
\item Max Morrison et al. “Controllable neural prosody synthesis”. In: arXiv preprint arXiv:2008.03388 (2020).
\item Moseli Mots’ oehli, Anna Sergeevna Bosman, and Johan Pieter De Villiers. “Comparision Of Adversarial And Non-Adversarial LSTM Music Generative Models”. In: arXiv preprint arXiv:2211.00731 (2022). 
\item Ahmed Mustafa et al. “Analysis by Adversarial Synthesis–A Novel Approach for Speech Vocoding”. In: arXiv preprint arXiv:1907.00772 (2019).
\item Zaha Mustafa Badi and Lamia Fathi Abusedra. “Neural Network-based Vocoders in Arabic Speech Synthesis”. In: The 7th International Conference on Engineering \& MIS 2021. 2021, pp. 1–5. 
\item Tomohiro Nakatani and Keisuke Kinoshita. “Maximum likelihood convolu- tional beamformer for simultaneous denoising and dereverberation”. In: 2019 27th European Signal Processing Conference (EUSIPCO). IEEE. 2019, pp. 1–5. 
\item Pedro Neves, Jose Fornari, and Jo\~ao Florindo. “Generating music with sentiment using Transformer-GANs”. In: arXiv preprint arXiv:2212.11134 (2022). 
\item Aditya Arie Nugraha, Kouhei Sekiguchi, and Kazuyoshi Yoshii. “A deep gen- erative model of speech complex spectrograms”. In: ICASSP 2019-2019 IEEE International Conference on Acoustics, Speech and Signal Processing (ICASSP). IEEE. 2019, pp. 905–909. 
\item Manuel Pariente, Antoine Deleforge, and Emmanuel Vincent. “A statistically principled and computationally efficient approach to speech enhancement using variational autoencoders”. In: arXiv preprint arXiv:1905.01209 (2019). 
\item Sangwook Park, David K Han, and Hanseok Ko. “Sinusoidal wave generating network based on adversarial learning and its application: synthesizing frog sounds for data augmentation”. In: arXiv preprint arXiv:1901.02050 (2019).
\item  Ashis Pati and Alexander Lerch. “Is disentanglement enough? On latent repre- sentations for controllable music generation”. In: arXiv preprint arXiv:2108.01450 (2021). 
\item Ashis Pati, Alexander Lerch, and Ga\"etan Hadjeres. “Learning to traverse latent spaces for musical score inpainting”. In: arXiv preprint arXiv:1907.01164 (2019). 
\item Adam Polyak et al. “High fidelity speech regeneration with application to speech enhancement”. In: ICASSP 2021-2021 IEEE International Conference on Acoustics, Speech and Signal Processing (ICASSP). IEEE. 2021, pp. 7143–7147. 
\item Adam Polyak et al. “Unsupervised cross-domain singing voice conversion”. In: arXiv preprint arXiv:2008.02830 (2020). 
\item Rohan Proctor and Charles Patrick Martin. “A Laptop Ensemble Performance System using Recurrent Neural Networks”. In: arXiv preprint arXiv:2012.02322 (2020).
\item Jie Pu, Yixiong Meng, and Oguz Elibol. “Building Synthetic Speaker Profiles in Text-to-Speech Systems”. In: arXiv preprint arXiv:2202.03125 (2022). 
\item Hendrik Purwins et al. “Deep learning for audio signal processing”. In: IEEE Journal of Selected Topics in Signal Processing 13.2 (2019), pp. 206–219. 
\item Shakeel Raja. “Music generation with temporal structure augmentation”. In: arXiv preprint arXiv:2004.10246 (2020). 
\item Yi Ren et al. “Popmag: Pop music accompaniment generation”. In: Proceedings of the 28th ACM international conference on multimedia. 2020, pp. 1198–1206. 
\item Julius Richter et al. “Speech enhancement and dereverberation with diffusion- based generative models”. In: arXiv preprint arXiv:2208.05830 (2022). 
\item Simon Rouard and Ga\"etan Hadjeres. “CRASH: raw audio score-based gener- ative modeling for controllable high-resolution drum sound synthesis”. In: arXiv preprint arXiv:2106.07431 (2021).
\item Yuki Saito, Shinnosuke Takamichi, and Hiroshi Saruwatari. “Perceptual-similarity- aware deep speaker representation learning for multi-speaker generative mod- eling”. In: IEEE/ACM Transactions on Audio, Speech, and Language Processing 29 (2021), pp. 1033–1048.
\item Ryosuke Sawata et al. “A Versatile Diffusion-based Generative Refiner for Speech Enhancement”. In: arXiv preprint arXiv:2210.17287 (2022).
\item Kouhei Sekiguchi et al. “Semi-supervised multichannel speech enhancement with a deep speech prior”. In: IEEE/ACM Transactions on Audio, Speech, and Language Processing 27.12 (2019), pp. 2197–2212.
\item Joan Serr\'a et al. “Universal speech enhancement with score-based diffusion”. In: arXiv preprint arXiv:2206.03065 (2022).
\item Ravi Shankar, Jacob Sager, and Archana Venkataraman. “Non-parallel emotion conversion using a deep-generative hybrid network and an adversarial pair discriminator”. In: arXiv preprint arXiv:2007.12932 (2020).
\item Berrak Sisman et al. “An Overview of Voice Conversion and Its Challenges: From Statistical Modeling to Deep Learning”. In: 29 (2020), 132–157. issn: 2329-9290. 
\item Eunwoo Song et al. Neural text-to-speech with a modeling-by-generation excita- tion vocoder. July 2020.
\item Qingwei Song et al. “SinTra: Learning an inspiration model from a single multi-track music segment”. In: arXiv preprint arXiv:2204.09917 (2022).
\item Krishna Subramani and Preeti Rao. “Hprnet: Incorporating residual noise modeling for violin in a variational parametric synthesizer”. In: arXiv preprint arXiv:2008.08405 (2020). 
\item Minhyang (Mia) Suh et al. “AI as Social Glue: Uncovering the Roles of Deep Generative AI during Social Music Composition”. In: Proceedings of the 2021 CHI Conference on Human Factors in Computing Systems. CHI ’21. Yokohama, Japan: Association for Computing Machinery, 2021. isbn: 9781450380966. 
\item Jaesung Tae, Hyeongju Kim, and Taesu Kim. “EdiTTS: Score-based editing for controllable text-to-speech”. In: arXiv preprint arXiv:2110.02584 (2021). 
\item Naoya Takahashi, Mayank Kumar, Yuki Mitsufuji, et al. “Hierarchical Diffusion Models for Singing Voice Neural Vocoder”. In: arXiv preprint arXiv:2210.07508 (2022). 
\item Hao Hao Tan, Yin-Jyun Luo, and Dorien Herremans. “Generative modelling for controllable audio synthesis of expressive piano performance”. In: arXiv preprint arXiv:2006.09833 (2020).
\item Vibert Thio et al. “A minimal template for interactive web-based demonstra- tions of musical machine learning”. In: arXiv preprint arXiv:1902.03722 (2019). 
\item Georgi Tinchev et al. “Modelling low-resource accents without accent-specific TTS frontend”. In: arXiv preprint arXiv:2301.04606 (2023). 
\item Maciej Tomczak, Masataka Goto, and Jason Hockman. “Drum synthesis and rhythmic transformation with adversarial autoencoders”. In: Proceedings of the 28th ACM International Conference on Multimedia. 2020, pp. 2427–2435. 
\item Se-Yun Um et al. “Facetron: A Multi-speaker Face-to-Speech Model based on Cross-modal Latent Representations”. In: arXiv preprint arXiv:2107.12003 (2021). 
\item Jean-Marc Valin et al. “Real-time packet loss concealment with mixed genera- tive and predictive model”. In: arXiv preprint arXiv:2205.05785 (2022).
\item Sean Vasquez and Mike Lewis. “Melnet: A generative model for audio in the frequency domain”. In: arXiv preprint arXiv:1906.01083 (2019). 
\item Prateek Verma and Chris Chafe. “A generative model for raw audio using transformer architectures”. In: 2021 24th International Conference on Digital Audio Effects (DAFx). IEEE. 2021, pp. 230–237.
\item Dominik Wagner et al. “Generative Models for Improved Naturalness, Intel- ligibility, and Voicing of Whispered Speech”. In: 2022 IEEE Spoken Language Technology Workshop (SLT). IEEE. 2023, pp. 943–948. 
\item Run Wang et al. “DeepSonar: Towards Effective and Robust Detection of AI- Synthesized Fake Voices”. In: Proceedings of the 28th ACM International Con- ference on Multimedia. MM ’20. Seattle, WA, USA: Association for Computing Machinery, 2020, 1207–1216. isbn: 9781450379885. 
\item Songhe Wang, Zheng Bao, and Jingtong E. “Armor: A Benchmark for Meta- evaluation of Artificial Music”. In: Proceedings of the 29th ACM International Conference on Multimedia. 2021, pp. 5583–5590.
\item Tao Wang et al. “Neuraldps: Neural deterministic plus stochastic model with multiband excitation for noise-controllable waveform generation”. In: IEEE/ACM Transactions on Audio, Speech, and Language Processing 30 (2022), pp. 865–878. 
\item Xin Wang, Shinji Takaki, and Junichi Yamagishi. “Neural source-filter-based waveform model for statistical parametric speech synthesis”. In: ICASSP 2019- 2019 IEEE International Conference on Acoustics, Speech and Signal Processing (ICASSP). IEEE. 2019, pp. 5916–5920. 
\item Xin Wang, Shinji Takaki, and Junichi Yamagishi. “Neural source-filter wave- form models for statistical parametric speech synthesis”. In: IEEE/ACM Trans- actions on Audio, Speech, and Language Processing 28 (2019), pp. 402–415.
\item  Ziyu Wang et al. “Learning interpretable representation for controllable poly- phonic music generation”. In: arXiv preprint arXiv:2008.07122 (2020). 
\item Ron J Weiss et al. “Wave-tacotron: Spectrogram-free end-to-end text-to-speech synthesis”. In: ICASSP 2021-2021 IEEE International Conference on Acoustics, Speech and Signal Processing (ICASSP). IEEE. 2021, pp. 5679–5683.
\item Simon Welker, Julius Richter, and Timo Gerkmann. “Speech enhancement with score-based generative models in the complex STFT domain”. In: arXiv preprint arXiv:2203.17004 (2022). 
\item Matt Whitehill et al. “Multi-reference neural TTS stylization with adversarial cycle consistency”. In: arXiv preprint arXiv:1910.11958 (2019).
\item William J Wilkinson, Joshua D Reiss, and Dan Stowell. “A generative model for natural sounds based on latent force modelling”. In: Latent Variable Analysis and Signal Separation: 14th International Conference, LVA/ICA 2018, Guildford, UK, July 2–5, 2018, Proceedings 14. Springer. 2018, pp. 259–269. 
\item Da-Yi Wu and Yi-Hsuan Yang. “Speech-to-singing conversion based on bound- ary equilibrium GAN”. In: arXiv preprint arXiv:2005.13835 (2020).
\item Da-Yi Wu et al. “DDSP-based singing vocoders: A new subtractive-based syn- thesizer and a comprehensive evaluation”. In: arXiv preprint arXiv:2208.04756 (2022). 
\item Guowei Wu, Shipei Liu, and Xiaoya Fan. The Power of Fragmentation: A Hi- erarchical Transformer Model for Structural Segmentation in Symbolic Music Generation. July 2022.
\item Shih-Lun Wu and Yi-Hsuan Yang. “The Jazz Transformer on the front line: Exploring the shortcomings of AI-composed music through quantitative mea- sures”. In: arXiv preprint arXiv:2008.01307 (2020). 
\item Yi-Chiao Wu et al. “Quasi-periodic parallel WaveGAN: a non-autoregressive raw waveform generative model with pitch-dependent dilated convolution neural network”. In: IEEE/ACM Transactions on Audio, Speech, and Language Processing 29 (2021), pp. 792–806. 
\item Yi-Chiao Wu et al. “Quasi-periodic WaveNet: An autoregressive raw waveform generative model with pitch-dependent dilated convolution neural network”. In: IEEE/ACM Transactions on Audio, Speech, and Language Processing 29 (2021), pp. 1134–1148. 
\item Yang Xiang and Changchun Bao. “A parallel-data-free speech enhancement method using multi-objective learning cycle-consistent generative adversarial network”. In: IEEE/ACM Transactions on Audio, Speech, and Language Processing 28 (2020), pp. 1826–1838. 
\item Yuying Xie, Thomas Arildsen, and Zheng-Hua Tan. “Complex Recurrent Varia- tional Autoencoder for Speech Enhancement”. In: arXiv preprint arXiv:2204.02195 (2022).
\item Heyang Xue et al. “Noise Robust Singing Voice Synthesis Using Gaussian Mixture Variational Autoencoder”. In: Companion Publication of the 2021 Inter- national Conference on Multimodal Interaction. 2021, pp. 131–136. 
\item Ryuichi Yamamoto, Eunwoo Song, and Jae-Min Kim. “Parallel WaveGAN: A fast waveform generation model based on generative adversarial networks with multi-resolution spectrogram”. In: ICASSP 2020-2020 IEEE International Conference on Acoustics, Speech and Signal Processing (ICASSP). IEEE. 2020, pp. 6199–6203. 
\item Hao Yen et al. “Cold Diffusion for Speech Enhancement”. In: arXiv preprint arXiv:2211.02527 (2022).
\item Takenori Yoshimura et al. “Mel-cepstrum-based quantization noise shaping applied to neural-network-based speech waveform synthesis”. In: IEEE/ACM Transactions on Audio, Speech, and Language Processing 26.7 (2018), pp. 1177– 1184. 
\item Chenyu You, Nuo Chen, and Yuexian Zou. “Self-supervised contrastive cross- modality representation learning for spoken question answering”. In: arXiv preprint arXiv:2109.03381 (2021).
\item Yi Yu, Abhishek Srivastava, and Simon Canales. “Conditional lstm-gan for melody generation from lyrics”. In: ACM Transactions on Multimedia Comput- ing, Communications, and Applications (TOMM) 17.1 (2021), pp. 1–20. 
\item Chen Zhang et al. “SDMuse: Stochastic Differential Music Editing and Genera- tion via Hybrid Representation”. In: arXiv preprint arXiv:2211.00222 (2022).
\item  Jianwei Zhang, Suren Jayasuriya, and Visar Berisha. “Restoring degraded speech via a modified diffusion model”. In: arXiv preprint arXiv:2104.11347 (2021). 
\item Kexun Zhang et al. “WSRGlow: A Glow-based waveform generative model for audio super-resolution”. In: arXiv preprint arXiv:2106.08507 (2021).
\item Lu Zhang et al. “Incorporating multi-target in multi-stage speech enhancement model for better generalization”. In: 2021 Asia-Pacific Signal and Information Processing Association Annual Summit and Conference (APSIPA ASC). IEEE. 2021, pp. 553–558. 
\item Jingwei Zhao and Gus Xia. “Accomontage: Accompaniment arrangement via phrase selection and style transfer”. In: arXiv preprint arXiv:2108.11213 (2021). 
\item Yizhou Zhao et al. “Vertical-Horizontal Structured Attention for Generating Music with Chords”. In: arXiv preprint arXiv:2011.09078 (2020). 
\item Ziyi Zhao et al. “A Review of Intelligent Music Generation Systems”. In: arXiv preprint arXiv:2211.09124 (2022).
\item Cong Zhou et al. “Voice conversion with conditional SampleRNN”. In: arXiv preprint arXiv:1808.08311 (2018). 
\item Yijun Zhou et al. “Generative melody composition with human-in-the-loop Bayesian optimization”. In: arXiv preprint arXiv:2010.03190 (2020). 
\item Hongyuan Zhu et al. “Pop music generation: From melody to multi-style arrangement”. In: ACM Transactions on Knowledge Discovery from Data (TKDD) 14.5 (2020), pp. 1–31. 
\item Guo Zixun, Dimos Makris, and Dorien Herremans. “Hierarchical recurrent neural networks for conditional melody generation with long-term structure”. In: 2021 International Joint Conference on Neural Networks (IJCNN). IEEE. 2021, pp. 1–8. 
\end{enumerate}
\end{document}